

\long\def\UN#1{$\underline{{\vphantom{\hbox{#1}}}\smash{\hbox{#1}}}$}
\def\NI{\noindent}

\magnification=\magstep 1
\overfullrule=0pt
\hfuzz=16pt
\voffset=0.0 true in
\vsize=8.8 true in
\baselineskip 12pt
\parskip 2pt
\hoffset=0.1 true in
\hsize=6.3 true in
\nopagenumbers
\pageno=1
\footline={\hfil -- {\folio} -- \hfil}

\

\NI{}TITLE of the contribution to the book\ \ {\sl Trends in
Statistical Physics:}

\

{{\bf Dynamics of Nonequilibrium Processes: Surface
Adsorption,}}

{{\bf Reaction-Diffusion Kinetics, Ordering and Phase
Separation}}

\

\

\NI{}AUTHOR and affiliation:

\

{\bf Vladimir Privman}

{\sl $\qquad$Department of Physics, Clarkson University,}

{\sl $\qquad$Potsdam, New York 13699--5820, USA}

\

\

\vskip 0.12 in

\NI \UN{\bf INTRODUCTION}

This short review covers a wide selection of topics from a
multidisciplinary area of dynamics of nonequilibrium systems in
physics, chemistry, biology. Theoretical models of colloid particle
and protein deposition and adhesion at surfaces, accompanied by
relaxation processes, of reaction kinetics with and without
diffusion, of phase coarsening and nucleation, will be surveyed. The
unifying feature of these systems is the importance of fluctuations
and many-body, multiparticle collective effects, in determining the
dynamical behavior which is typically nonclassical.

Recently a range of
methods has been employed to study models of these
phenomena, including exact solutions, analytical and
asymptotic methods, and large-scale numerical simulations. The
review is largely devoted to summary of several recent advances and
provides a guide to the literature. No detailed derivations are
given; our aim is to overview the field emphasizing the
unifying aspects of various phenomena surveyed and list some open
research directions.

Most actual applications of the methods of statistical physics to
practical problems as well as to scientific experiment
interpretation involve time-dependent processes. Generally,
nonequilibrium statistical physics has developed slower than
equilibrium theories. However, recently there has been a
resurgence of interest and activity, as well as reports of new
advances, in the field of nonequilibrium statistical mechanical
systems.

The reason for this recent progress has been several-fold. Firstly,
guided by experience with the equilibrium case, attention was
focussed on lattice models with stochastic dynamics, rather than on
direct examination of phenomenological, continuum evolution
equations popular in ``traditional'' studies of reaction-diffusion
and nucleation phenomena. Secondly, emphasis has been put on
identifying those regimes where the system's behavior is
dominated by nonclassical, non-mean-field fluctuations. This has
lead to consideration of rather low dimensions, one (chains) and two
(surfaces).

Finally, theoretical advances have been
stimulated by the fact that these low-dimensional models were
actually relevant to experimental studies of surface deposition of
colloidal particles and proteins, heterogeneous catalysis, reaction
kinetics of excitations in polymer chains, interactions of
biological molecules such as DNA with small molecules, etc.
This new interdisciplinary horizon of applications has boosted
research and brought in ideas, nomenclature, and emphases from a
diverse scientific background in biology, chemistry, physics.

\vskip 0.12in

\NI \UN{\bf SURFACE DEPOSITION}

\NI \UN{\sl Random Sequential Adsorption Models}

Random sequential adsorption (RSA) models have been studied
extensively due to their relevance to deposition processes on
surfaces (reviews by Bartelt \& Privman 1991, Evans 1993). The
depositing particles are represented by hard-core extended objects;
they are not allowed to overlap. In monolayer deposition of
colloidal particles and macromolecules (Feder \& Giaever 1980,
Onoda \& Liniger 1986, Ryde et al.~1991, 1992, Song \& Elimelech
1993) one can further
assume that the adhesion process is irreversible.

However, recent experiments on
protein adhesion at surfaces (Ramsden 1992, 1993) indicate that in
biomolecular systems effects of surface relaxation, due to
diffusional rearrangement of particles, are observable on time
scales of the deposition process. The resulting large-time coverage
is denser than in irreversible RSA and in fact it is experimentally
comparable to the fully packed (i.e., locally semi-crystalline)
particle arrangement.

Irreversible RSA has been studied extensively by many authors
(reviews by Bartelt \& Privman 1991, Evans 1993). The most
interesting aspect of such processes is the power-law large-time
convergence to the jamming coverage in continuum off-lattice
deposition. This slow time-dependence, as opposed to exponential
convergence in lattice deposition, is due to gaps arbitrary close
in size (and shape) to that of the depositing particles and
therefore reached with low probability. Asymptotic arguments
describe this rare-event dominated process (Pomeau 1980, Swendsen
1981). Crossover from lattice to continuum can be also elucidated
analytically (Privman et al.~1991b).

\NI \UN{\it Diffusional and Detachment Relaxation}

Studies of RSA with diffusional relaxation by
analytical means encounter several difficulties associated with
collective effects in hard-core particle systems at high
densities (such as, for instance, phase separation), and with the
possibility, in certain lattice models, of locally ``gridlocked''
vacant sites. The latter effect may actually prevent full coverage
in some models; this matter remains an open problem at this time.

Both difficulties are not present in $1d$: there are no
equilibrium phase transitions (in models without deposition), traces
of which might manifest themselves as collective effects in $d>1$
deposition with diffusion (Wang et al.~1993a), and furthermore
diffusional relaxation leads to simple hopping-diffusion
interpretation of the motion of vacant sites in $1d$ which
recombine to form larger open voids accessible to deposition
attempts. As a result, both accurate numerical studies and their
analytical interpretation have been possible in $1d$ (Nielaba \&
Privman 1992a, Privman \& Barma 1992, Privman \& Nielaba 1992).

For $d>1$ models, a low-density-expansion approximation scheme was
applied (Tarjus et al. 1990) to off-lattice deposition of circles on
a plane, accompanied by diffusional relaxation. However, no
analytical studies of the high-density behavior and the associated
collective effects, were reported for circles.

Recently, rather extensive numerical simulations were reported (Wang
et al.~1993a), of the RSA process with diffusional relaxation, for
the $2d$ lattice hard-square model (review by Runnels 1972), i.e.,
the square-lattice hard-core model with nearest neighbor exclusion.
This model is well studied for its equilibrium phase transition
(review by Runnels 1972) which is second-order with disordered phase at low
densities and two coexisting ordered phases, corresponding to two
different sublattice particle coverage arrangements at high
densities. Another simplifying feature of the hard-square model is
that the only possible gridlocked (locally frozen) vacancies are
parts of domain walls (Wang et
al.~1993a). As a result the coverage reaches the full
crystalline limit at large times, by a process of diffusional
domain wall motion leading to cluster growth reminiscent of
quenched binary alloys and fluids at low temperatures (Gunton et
al.~1983, Mouritsen 1990, Sadiq \& Binder 1984).

For $2 \times 2$ lattice squares, extensive numerical results were
reported as well (Wang et al.~1993b). However, their
interpretation is still incomplete. Indeed, $2 \times 2$
lattice squares form for large times and on periodic lattices of
even dimensions, a state with frozen single-site gridlocked
vacancies. Diffusional interpretation of the relaxation processes
is less apparent, and many open questions remain to be answered by
future more extensive numerical studies; see Wang et al.~(1993b) for
details.

Relaxation by detachment is much more difficult to quantify
experimentally, at least in colloidal systems, because there are
many types of adhesion states of particles, each detaching
differently. However, some interesting theoretical aspects present
themselves to investigation. Indeed, similar to diffusional
relaxation the late stages will be dominated by recombination of
smaller empty area regions into larger ones which can accommodate
arriving particles. Power-law tails are expected in some regimes.
Systematic theoretical studies are very recent (Chen et al.~1993,
Dhar \& Barma 1993, Krapivsky \& Ben-Naim 1993); the field is
still widely open.

\NI \UN{\it Multilayer Deposition}

Multilayer deposit formation is a rather wide field of research
with most theoretical activity recently focussed on growing
interface fluctuations. However, in colloid deposition experiments
(Ryde et al.~1992, Song \& Elimelech
1993; see also Ryde et al.~1991) one is frequently
interested in the regime where the amount deposited is limited, and
the dynamics of the formation of amorphous deposits is RSA-like,
i.e., governed by a combination of exclusion effects (due to
particle size) and screening. Theoretical progress has proved
difficult and the available results were either mean-field theories
(Privman et al.~1991a) which in fact were applied to
the experimental data analyses (Ryde et al.~1992, Song \& Elimelech
1993), or large-scale numerical simulations (Bartelt \& Privman
1990, Lubachevsky
et al.~1993, Nielaba \& Privman 1992b, Nielaba et al.~1990).

Specifically, numerical simulations of deposition on $2d$ substrates
(as well as model $1d$ substrates) provide data on the deposit
density and morphology as it varies from the wall-induced density
oscillations (Lubachevsky et al.~1993) to the bulk,
interface-fluctuation dominated power-law behavior (Jullien \&
Meakin 1987, Krug 1989, Krug \& Meakin 1990). Much work remains to
be done; the main challenge is the numerical effort needed for such
simulations.

\vskip 0.12in

\NI {\bf REACTION KINETICS}

\NI \UN{\it Low-Dimensional Models and Fluctuations}

Reaction-diffusion systems in low dimensions have been
investigated extensively recently with emphasis on
fluctuation-dominated effects, specifically, the breakdown of the
standard chemical rate equations which correspond to the
``mean-field'' approximation of the reaction kinetics. Recent
interest has been largely focused on the simplest reactions of
two-particle coagulation, $A+A \to A$, and annihilation,
$A+A \to {\rm inert}$, on the $1d$ lattice (Amar \& Family
1990, Balding \& Green 1989, ben-Avraham et al.~1990, Bramson \&
Griffeath 1980, Bramson \& Lebowitz 1988, Doering \& ben-Avraham
1988, Kang et al.~1984, Kuzovkov \& Kotomin 1988, Liggett 1985,
Lin et al.~1990, Lushnikov 1987, Privman 1992a, 1993a,b, Racz
1985, Spouge 1988, Torney \& McConnell 1983). Indeed, in the
diffusion-limited, instantaneous reaction case, these processes
show non-mean-field power-law decay of the $A$-particle density.
Experimental verification of the theoretical predictions has been
initiated recently (Kopelman et al.~1990, Kroon et al.~1993).

Exact
solutions provide convenient reference and guide to more
complicated and more realistic systems. As usual in other fields,
the number of exactly solvable cases is limited both in
dimensionality, to $d=1$, and in the variety of models that can be
solved. Thus, future progress in this field will be largely based on
approximation techniques such as diffusive description of the
interparticle distribution, elaborate mean-field approximations,
and extensive numerical Monte Carlo simulations.

\NI \UN{\it Partial-Probability Reactions}

Experimentally, reactions (Kopelman et
al. 1990, Kroon et al.~1993) are never instantaneous. There is
always some probability for the particles, identical $A+A$ or
different-species $A+B$, to bounce off (modeled, e.g., by the
hard-core repulsion) rather than interact to yield the reaction
products. While the first numerical results for the simplest,
$(A+A)$-type $1d$ reactions have been reported (Braunstein et
al.~1992, Martin \& Braunstein 1993, Shi \& Kopelman 1992a,b)
recently, theoretical progress has been achieved only for the $1d$
coagulation-reaction case (Privman et al.~1993) by an approximate
model description of the interparticle distance distribution within
a diffusive scheme in which the partial reaction is modeled
phenomenologically by the radiation boundary conditions while the
hard-core is modeled by a certain drift term in the diffusion
current.

Generally, the field is still wide open; numerical studies and
analytical efforts are needed, in $d>1$ as well as for more
complicated reactions, specifically, $(A+B)$-type, reversible
reactions, and particle-input reactions. In the latter cases, there
is a steady state which may be easier to handle in the
diffusive-approximation schemes (Nielaba \& Privman 1992a,
Privman \& Barma 1992, Privman et al.~1993, Privman \& Grynberg
1992, Privman \& Nielaba 1992) which are typically
nonlinear in time but linear in the spatial coordinate
dependence. The analytical model building is slow and requires good
physical insight into the underlying reaction process. Thus,
extensive numerical data generation and interpretation will be
essential to ensure progress.

\NI \UN{\it Fast-Diffusion Mean-Field Theories}

Consideration of reactions with more
than two particles in the input, for
instance $kA \to  {\rm inert}$, etc., leads to use of mean-field
theories in $d=1,2$. Indeed, such
reactions are asymptotically mean-field
even in $1d$, for $k>3$. The marginal case $k=3$ should have
logarithmic corrections to the rate-equation behavior. These were
observed numerically recently (see ben-Avraham 1993), and also
derived by field-theoretical methods (Lee 1993). The two-particle
reactions are marginal in $d=2$ or higher.

Study of multiparticle reactions has been emphasized
recently due to relevance to certain deposition processes (Nielaba
\& Privman 1992a, Privman \& Barma 1992, Privman \& Grynberg
1992, Privman \& Nielaba 1992). The usual mean-field approximations
involve closures of hierarchies of relations for correlation
functions. However, in $d=1$ a different, ``fast-diffusion''
mean-field approach is possible based on the use of the
interparticle distance distribution (which is in a sense a
complicated correlation function) from the fast-diffusion limit.
The advantage of this approach is that it allows derivation of
criteria of the applicability of the mean-field theory (Privman \&
Grynberg 1992) for
reaction-diffusion systems, reminiscent of the Ginzburg criteria in
static critical phenomena. An open challenge is extension to
$d>1$: for large densities of particles, in $d>1$, fast
diffusion can lead to collective effects, as mentioned earlier.
Accommodation of these within the mean-field approach has not been
clarified thus far.

\NI \UN{\it Immobile Reactants}

Another solvable limit, in $1d$ (Kenkre \& Van Horn 1981),
is the case of no diffusion at all. Generally such models of
immobile reactants  have received less attention
in the literature (Kenkre \& Van Horn 1981, Schn\" orer et
al.~1989, 1990, Majumdar \& Privman 1993). The reason is that
unless longer-range reactions are included (Schn\" orer et
al.~1989, 1990), the time
dependence involves exponential relaxation to an absorbing state
rather than power-law behavior typical of the fast diffusion
reactions. Thus there are no universal fluctuation effects involved.

On the other hand, immobile-reactant systems provide an
example of freezing in an absorbing state with a
nonuniversal, initial-condition-dependent behavior
persistent at all times and, again, not consistent with
mean-field rate equations. It is therefore of interest to
derive exact results whenever possible. In a recent work (Majumdar
\& Privman 1993) we  reported an exact solution for $A+A \to {\rm
inert}$ on the Bethe lattice. In $d=1$ we also derived exact
results for the reaction $A+B \to {\rm inert}$, extending the
previously know solution of $A+A \to {\rm inert}$ (Kenkre \& Van
Horn 1981). The techniques used are similar to RSA and to certain
models surveyed in the section on ordering, below. Thus there are
further cases to be studied, specifically, multiparticle-input and
systems of more than two species.

\vskip 0.12in

\NI {\bf ORDERING AND NUCLEATION}

\NI \UN{\it Lattice Models of Phase Separation}

Recently there has been much interest in modeling phase separation
and spinodal decomposition (Gunton et
al.~1983, Mouritsen 1990, Sadiq \& Binder 1984) by simple
irreversible, effectively zero-temperature low dimensional
stochastic dynamical systems (Hede \& Privman 1991, Privman
1992a, Scheucher \& Spohn 1988). Specifically, some variants of
nonconserved order parameter dynamical models in $d=1$,
corresponding effectively to $T=0$ Glauber-type spin systems, have
been solved exactly for properties such as the structure factor and
average domain size, as functions of time; see Bray (1990), Privman
(1992a) for details. The underlying mechanism leading to cluster
growth in $d=1$ is pairwise annihilation of interfaces separating
ordered domains. The interfacial motion is diffusional and it
corresponds also to diffusion-limited particle annihilation models
discussed earlier.

Recently, pairwise particle-exchange models on linear and
Bethe lattices were solved exactly by a new rate-equation method
(Krapivsky 1993, Lin \& Taylor 1993, Majumdar \& Sire 1993,
Privman 1992b). Lattice sites are occupied by particles A and B
which can exchange irreversibly provided the local energy in
reduced. Thus, the model corresponds to a zero-temperature
Kawasaki-type phase separation process. Due to local order
parameter conservation, the dynamics reaches an absorbing, frozen
state at large times, the structure of which depends on the initial
conditions.

\NI \UN{\it Nonconserved Order Parameter Models}

The $T \to 0$ limiting model with Glauber-type dynamics involves
interface annihilation which is a process lowering the local energy
and therefore has Boltzmann factor $+\infty$ associated with its
transition probability at $T = 0$. Interface diffusion does not
change the local energy and therefore has Boltzmann factor $1$.
Finally, interface generation (birth) has Boltzmann factor $0$ (due
to energy cost) at $T=0$. The $T=0$ models referred to earlier,
correspond to allowing for both annihilation and diffusion. Exact
results for such models in $1d$ yield the structure factor
(Bray 1990, Privman 1992a).

In order to model nonsymmetric growth of a stable phase from an
unstable-phase (or mixed) initial state, the above ordering
mechanism was supplemented by
spontaneous creation of the stable phase, +1, regions by
overturning the unstable phase, $-1$, spins with probability $p$
(Privman 1992a). For cluster coarsening at phase coexistence,
$p=0$, the conventional structure-factor scaling applies. The $\pm
1$ cluster sizes grow diffusively, $\sim \sqrt{t}$, and the
two-point correlation function obeys scaling. However, for $p>0$,
i.e., for the dynamics of formation of stable phase from unstable
phase, the structure-factor scaling breaks down; the length
scale associated with the size of the growing $+1$ clusters
reflects only the short-distance properties of the two-point
correlations.

\NI \UN{\it Conserved Order Parameter Models}

For conserved order parameter, spin-exchange Kawasaki-type
dynamical models, there are several new features as compared to the
nonconserved models just surveyed. Notably, interfacial processes
even at $T=0$ are more complicated for the conserved case.
Specifically, let us consider the $d=1$ binary AB-mixture model:
each site of the $1d$ lattice is occupied by particle A or particle
B. The dynamics generally involves nearby particle exchanges; the
locally conserved order parameter is the difference of the A-\ and
B-particle densities.

Energy-conserving interfacial motion in $1d$ is no longer simple
free diffusion (Privman 1992b). Freezing rather than full phase
separation occurs asymptotically for large times in models with both
energy-lowering and energy-conserving moves allowed, or with only
energy-lowering moves allowed.
Several numerical studies were reported (Alcaraz et al.~1986, Levy
et al.~1982, Meakin \& Reich 1982, Palmer \& Frisch 1985) of such
particle-exchange models for $d$ up to 5. As in the nonconserved
case, some of the properties of the $d=1$ models are different from
$d>1$. However, the general expectation of the ``freezing'' of the
domain structure at large times applies, for conserved dynamics, in
all space dimensions.

\vskip 0.12in

\NI {\bf REFERENCES}{\frenchspacing

\NI\hang{}Alcaraz, F.C., Drugowich de
Fel\'icio, J.R. \& K\"oberle, R. \ 1986. \ {\sl Phys.
Lett.} {\bf 118}A, 200.

\NI\hang{}Amar, J.G. \& Family, F. \ 1990. \ {\sl
Phys. Rev.} A{\bf 41}, 3258.

\NI\hang{}Balding, D.J. \& Green, N.J.B. \
1989. \ {\sl Phys. Rev.} A{\bf 40}, 4585.

\NI\hang{}Bartelt, M.C. \& Privman, V. \
1990. \ {\sl J. Chem. Phys.} {\bf 93}, 6820.

\NI\hang{}Bartelt, M.C. \& Privman, V.
\ 1991. \ {\sl  Int.
J. Mod. Phys.} B{\bf 5}, 2883.

\NI\hang{}ben-Avraham, D. \ 1993. \  Preprint.

\NI\hang{}ben-Avraham, D., Burschka, M.A. \& Doering, C.R.
\ 1990. \ {\sl J. Stat. Phys.} {\bf 60}, 695.

\NI\hang{}Bramson, M. \& Griffeath, D.
\ 1980. \  {\sl Ann. Prob.} {\bf 8}, 183.

\NI\hang{}Bramson, M. \& Lebowitz, J.L.
\ 1988. \ {\sl Phys. Rev.
Lett.} {\bf 61}, 2397.

\NI\hang{}Braunstein, L., Martin, H.O., Grynberg, M.D. \&
Roman, H.E. \ 1992. \ {\sl J. Phys.} A{\bf 25}, L255.

\NI\hang{}Bray, A.J. \ 1990. \ {\sl J. Phys.} A{\bf 23}, L67.

\NI\hang{}Chen, N.N., Grynberg, M.D. \& Stinchcombe, R.B.
\ 1993. \ Preprint.

\NI\hang{}Dhar, D. \& Barma, M.
\ 1993. \ {\sl Pramana\/}
{\bf 41}, L193.

\NI\hang{}Doering, C.R. \& ben-Avraham, D.
\ 1988. \ {\sl Phys. Rev.}
A{\bf 38}, 3035.

\NI\hang{}Evans, J.W. \ 1993. \ {\sl Rev. Mod. Phys.}, in
print.

\NI\hang{}Feder, J. \& Giaever, I.
\ 1980. \ {\sl J. Colloid Interface
Sci.} {\bf 78}, 144.

\NI\hang{}Gunton, J.D., San Miguel, M. \& Sahni, P.S. \ 1983. \
In: Domb, C. \& Lebowitz, J.L. (Eds.) {\sl Phase Transitions and
Critical Phenomena}, Academic, London, Vol. 8, p. 267.

\NI\hang{}Hede, B. \& Privman, V.
\ 1991. \ {\sl J. Stat. Phys.} {\bf
65}, 379.

\NI\hang{}Jullien, R. \& Meakin, P.
\ 1987. \ {\sl Europhys. Lett.} {\bf 4}, 1385.

\NI\hang{}Kang, K., Meakin, P., Oh, J.H. \& Redner, S.
\ 1984. \ {\sl J. Phys.} A{\bf 17}, L665.

\NI\hang{}Kenkre, V.M. \& Van Horn, H.M.
\ 1981. \ {\sl Phys. Rev.} A{\bf 23}, 3200.

\NI\hang{}Kopelman, R., Li, C.S. \& Shi, Z.-Y.
\ 1990. \ {\sl J.
Luminescence\/} {\bf 45}, 40.

\NI\hang{}Krapivsky, P.L. \ 1993.  \ {\sl J. Stat. Phys.}, in
print.

\NI\hang{}Krapivsky, P.L. \& Ben-Naim, E. \ 1993. \ Preprint.

\NI\hang{}Kroon, R., Fleurent, H. \& Sprik, R.
\ 1993. \ {\sl Phys.
Rev.} E, in print.

\NI\hang{}Krug, J. \ 1989.
 \  {\sl J. Phys.} A{\bf 22}, L769.

\NI\hang{}Krug, J. \& Meakin, P.
\ 1990. \ {\sl J. Phys.} A{\bf 23}, L987.

\NI\hang{}Kuzovkov, V. \& Kotomin, E.
\ 1988. \ {\sl Rep. Prog. Phys.}
{\bf 51}, 1479.

\NI\hang{}Lee, B.P., \ 1993.  \  Preprint.

\NI\hang{}Levy, A., Reich, S. \& Meakin, P.
\ 1982. \ {\sl Phys. Lett.}
{\bf 87}A, 248.

\NI\hang{}Liggett, T. \ 1985. \ {\sl Interacting Particle
Systems}, \ Springer-Verlag, New York.

\NI\hang{}Lin, J.C., Doering, C.R.
\& ben-Avraham, D.  \ 1990. \ {\sl Chem.
Phys.} {\bf 146}, 355.

\NI\hang{}Lin, J.C. \&
Taylor, P.L. \ 1993.  \ Preprint.

\NI\hang{}Lubachevsky, B.D., Privman, V.
\& \ Roy, S.C. \ 1993. \ {\sl
Phys. Rev.} E{\bf 47}, 48.

\NI\hang{}Lushnikov, A.A. \ 1987.  \  {\sl Phys. Lett.} A{\bf
120}, 135.

\NI\hang{}Majumdar, S.N. \& Privman, V.
\ 1993. \ {\sl J.
Phys.} A{\bf 26}, L743.

\NI\hang{}Majumdar, S.N. \& Sire, C.
\ 1993. \ {\sl Phys.
Rev. Lett.} {\bf 70}, 4022.

\NI\hang{}Martin, H.O. \& Braunstein, L.
\ 1993. \ {\sl J.
Phys.} A, in print.

\NI\hang{}Meakin, P. \& Reich, S.
\ 1982. \ {\sl Phys. Lett.} {\bf
92}A, 247.

\NI\hang{}Mouritsen, O.G. \ 1990. \ In: Lagally, M.G.
(Ed.) {\sl Kinetics of Ordering and Growth at Surfaces},
Plenum, New York, p.~1.

\NI\hang{}Nielaba, P. \& Privman, V.
\ 1992a. \ {\sl Mod. Phys. Lett.}
B{\bf 6}, 533.

\NI\hang{}Nielaba, P. \& Privman, V.
\ 1992b. \ {\sl
Phys. Rev.} A{\bf 45}, 6099.

\NI\hang{}Nielaba, P., Privman, V. \&  Wang, J.-S.
\ 1990. \ {\sl J. Phys.} A{\bf 23}, L1187.

\NI\hang{}Onoda, G.Y. \& Liniger, E.G.
\ 1986. \ {\sl Phys. Rev.} A{\bf
33}, 715.

\NI\hang{}Palmer, R.G. \& Frisch, H.L.
\ 1985. \ {\sl J. Stat. Phys.}
{\bf 38}, 867.

\NI\hang{}Pomeau, Y. \ 1980.  \ {\sl J. Phys.} A{\bf 13},
L193.

\NI\hang{}Privman, V. \ 1992a.  \ {\sl J. Stat. Phys.} {\bf 69},
629.

\NI\hang{}Privman, V. \ 1992b.  \ {\sl Phys. Rev. Lett.} {\bf 69},
3686.

\NI\hang{}Privman, V.
\ 1993a.  \ {\sl J. Stat. Phys.} {\bf 72},
845.

\NI\hang{}Privman, V. \ 1993b.  \ Preprint.

\NI\hang{}Privman, V. \& Barma, M.
\ 1992. \ {\sl J. Chem. Phys.} {\bf
97}, 6714.

\NI\hang{}Privman, V., Doering, C.R. \& Frisch, H.L.
\ 1993. \ {\sl Phys. Rev.} E{\bf 48}, 846.

\NI\hang{}Privman, V.,
Frisch, H.L., Ryde, N. \& Matijevi\'c,
E. \ 1991a.  \ {\sl J. Chem. Soc. Farad. Tr.}
{\bf 87}, 1371.

\NI\hang{}Privman, V. \& Grynberg, M.D.
\ 1992. \ {\sl J. Phys.} A{\bf
25}, 6567.

\NI\hang{}Privman, V. \& Nielaba, P.
\ 1992. \ {\sl Europhys. Lett.}
{\bf 18}, 673.

\NI\hang{}Privman, V., Wang, J.-S. \& Nielaba, P.
\ 1991b. \ {\sl
Phys. Rev.} B{\bf 43}, 3366.

\NI\hang{}Racz, Z. \ 1985.  \ {\sl Phys. Rev. Lett.} {\bf 55},
1707.

\NI\hang{}Ramsden, J.J. \ 1992.  \ {\sl J. Phys. Chem.} {\bf
96}, 3388.

\NI\hang{}Ramsden, J.J. \ 1993.  \ {\sl J. Stat. Phys.}, in
print.

\NI\hang{}Runnels, L.K. \ 1972. \ In:
Domb, C. \& Green, M.S. (Eds.) {\sl Phase Transitions and
Critical Phenomena}, Academic, London, Vol. 2, p. 305.

\NI\hang{}Ryde, N., Kallay, N. \& Matijevi\'c, E.
\ 1991. \ {\sl J. Chem. Soc. Farad. Tr.} {\bf
87}, 1377.

\NI\hang{}Ryde, N., Kihira, H. \& Matijevi\'c, E.
\ 1992. \ {\sl J. Colloid Interface Sci.} {\bf
151}, 421.

\NI\hang{}Sadiq, A. \& Binder, K.
\ 1984. \ {\sl J. Stat. Phys.} {\bf
35}, 517.

\NI\hang{}Scheucher, M. \& Spohn, H.
\ 1988. \ {\sl J. Stat. Phys.} {\bf
53}, 279.

\NI\hang{}Schn\"orer, H., Kuzovkov, V. \& Blumen, A.
\ 1989. \ {\sl Phys. Rev. Lett.} {\bf 63}, 805.

\NI\hang{}Schn\"orer, H., Kuzovkov, V. \& Blumen, A. \ 1990.\
{\sl J. Chem. Phys.} {\bf 92}, 2310.

\NI\hang{}Shi, Z.-Y. \& Kopelman, R.
\ 1992a. \ {\sl Chem. Phys.} {\bf 167}, 149.

\NI\hang{}Shi, Z.-Y. \& Kopelman, R.
\ 1992b. \ {\sl J. Phys. Chem.} {\bf 96}, 6858.

\NI\hang{}Song, L. \& Elimelech, M.
\ 1993. \ {\sl Colloids and Surfaces\/} {\bf 73}, 49.

\NI\hang{}Spouge, J.L. \ 1988.  \ {\sl Phys. Rev. Lett.} {\bf
60}, 871.

\NI\hang{}Swendsen, R.H. \ 1981.  \ {\sl Phys. Rev.} A{\bf
24}, 504.

\NI\hang{}Tarjus, G., Schaaf, P. \& Talbot, J.
\ 1990. \ {\sl J. Chem.
Phys.} {\bf 93}, 8352.

\NI\hang{}Torney, D.C. \& McConnell, H.M.
\ 1983. \ {\sl J. Phys. Chem.} {\bf 87}, 1941.

\NI\hang{}Wang, J.-S., Nielaba, P. \& Privman, V.
\ 1993a. \ {\sl Mod.
Phys. Lett.} B{\bf 7}, 189.

\NI\hang{}Wang, J.-S., Nielaba, P. \& Privman, V.
\ 1993b. \ {\sl
Physica\/} A{\bf 199}, 527.

}\bye